\documentclass{emulateapj}

\newcommand{\um}{$\mu$m~}
\newcommand{\ums}{$\mu$m}

\shorttitle{SMART}
\shortauthors{Higdon et al.}

\begin{document}


\title{The SMART Data Analysis Package for the Infrared
  Spectrograph\footnotemark [1] on the Spitzer Space Telescope
  \footnotemark [2]}


\author{S. J. U. Higdon\altaffilmark{3}, 
D. Devost\altaffilmark{3}, J. L. Higdon\altaffilmark{3}, 
B. R. Brandl\altaffilmark{4}, J. R. Houck\altaffilmark{3}
P. Hall\altaffilmark{3}, D. Barry\altaffilmark{3}, 
V. Charmandaris\altaffilmark{3,5}, J. D. T. Smith\altaffilmark{6},
G. C. Sloan\altaffilmark{3}, \& J. Green\altaffilmark{7}}

\footnotetext [1] {The IRS was a
       collaborative venture between Cornell University and Ball
       Aerospace Corporation funded by NASA through the Jet Propulsion
       Laboratory and the Ames Research Center.}
\footnotetext [2] {The Spitzer Space Telescope
       is operated by JPL, California Institute of Technology for the
       National Aeronautics and Space Administration.}

\altaffiltext{3}{Astronomy Department, Cornell University, Ithaca, NY 14853; sjuh@isc.astro.cornell.edu}
\altaffiltext{4}{Leiden Observatory, 2300 RA Leiden, The Netherlands}
\altaffiltext{5}{Chercheur Associ\'e, Observatoire de Paris, F-75014, 
Paris, France}
\altaffiltext{6}{Steward Observatory, University of Arizona, Tucson, 
Arizona, 85721}
\altaffiltext{7}{Department of Physics and Astronomy, University of 
Rochester, Rochester, NY 14627}
%%University, Ithaca, NY 14853} 
\email{sjuh@astro.cornell.edu}
%\author{Sarah J. U. Higdon}%\altaffilmark{1,2,3} and \altaffilmark{1}}
%\affil{Cornell University,Ithaca NY 14853}
%\email{sjuh@astro.cornell.edu}
%\author{James L. Higdon}
%\affil{Cornell University,Ithaca NY 14853}
%\author{B. Brandl}
%\affil{Leiden University}

\newpage

\begin{abstract}
  
  SMART is a software package written in IDL to reduce and analyze
  Spitzer data from all four modules of the Infrared Spectrograph,
  including the peak-up arrays. The software is designed to make full
  use of the ancillary files generated in the Spitzer Science Center
  pipeline so that it can either remove or flag artifacts and
  corrupted data and maximize the signal-to-noise in the extraction
  routines. It may be run in both interactive and batch mode. The software
   and Users Guide will be available for public release in December
  2004. We briefly describe some of the main features of SMART
  including: visualization tools for assessing the data quality, basic
  arithmetic operations for either 2-d images or 1-d
  spectra, extraction of both point and extended sources and a suite
  of spectral analysis tools.

\end{abstract}


\newpage

\section{Introduction}

The Spectroscopy Modeling Analysis and Reduction Tool (SMART) is a
software package written in IDL \footnotemark [8] for the analysis of
data acquired with the Infrared Spectrograph\footnotemark [1] (IRS) on
the Spitzer Space Telescope \footnotemark [2] \citep{wer04}. The code
has been developed for the Unix/Linux operating systems. The IRS
comprises four separate spectrograph modules covering the wavelength
range from 5.3 to 38 \um with spectral resolutions, R $=\lambda
/\Delta \lambda \sim 90$ and 600.  The modules are named after their
wavelength coverage and resolution as Short-Low (SL), Short-High (SH),
Long-Low (LL) and Long-High(LH). The SL includes two peak-up imaging
cameras that have band-passes centered at 16 \um(``blue'') and 22
\um(``red''). For details of the IRS instrument see \citet{hou04} and
chapter 7 of the Spitzer Observers Manual\footnotemark [9] (SOM7).

\footnotetext [8] {The Interactive Data Language, Research Systems, Inc.}
\footnotetext [9] {\url{http://ssc.spitzer.caltech.edu/documents/som/}}

SMART has been designed specifically for IRS data and in particular to
extract spectra from observations of faint or extended sources. It has
been  written with an understanding of both the available Spitzer IRS
observing modes and a knowledge of how the contents of various files
generated by the Spitzer Science Center (SSC) IRS pipeline can be used
to maximize the signal-to-noise in the extracted spectrum. These three
design factors make it a comprehensive and powerful software package
for the extraction and analysis of IRS data.

SMART is primarily intended to operate on the basic calibrated data
(BCD, see SOM7) delivered by the SSC pipeline, but will also operate
on the browse quality data (BQD, including both images and wavelength
and flux calibrated spectral tables) and 2-d data products from
intermediate stages of the SSC pipeline, for example, the
un-flatfielded data. SMART aims to provide the routines necessary for
the processing and scientific analysis of IRS data. The main goal is
to simplify the tasks of visualizing, organizing, optimally
combining and extracting data. The result of this processing are fully
flux and wavelength calibrated spectra. Further analysis is available
within SMART.  Additionally, the spectra can be easily exported (in
either FITS, ascii or IDL save set format) to other analysis packages
written, for example, in IDL or IRAF.

SMART includes software developed by two of the Spitzer Legacy teams.
The Molecular Cores to Planet-Forming Disks (C2D) team has developed a
code to remove fringes caused by interference in the detector
substrate material \citep{fred02}.  This software is an enhanced
version of the code developed for the Infrared Space Observatory (ISO)
Short Wavelength Spectrometer \citep{kes03}.  The Formation and
Evolution of Planetary Systems (FEPS) team have adapted the
IDP3-NICMOS package \citep{stob02} to analyze image data from the IRS
peak-up cameras. The spectral analysis code is based on the inherited
ISO Spectral Analysis Package (ISAP\footnotemark [10] \footnotetext
[10] {The ISO Spectral Analysis Package (ISAP) is a joint development
  by the LWS and SWS Instrument Teams and Data Centers.  Contributing
  institutes are CESR, IAS, IPAC, MPE, RAL and SRON.})\citep{sturm98}.
The ISAP software is available at
\url{http://www.ipac.caltech.edu/iso/isap/isap.html}

The present paper is as an introduction to SMART. A SMART web
site at \url{http://isc.astro.cornell.edu/smart/} serves as the repository
for the full listing of all functions available in SMART and details
of the algorithms used. This website includes a comprehensive SMART
Users Guide (SUG) and a set of data reduction recipes aimed at the new user.
Each recipe outlines the steps required to produce wavelength and
flux calibrated spectra. Both the website and the software will
be publicly available in December 2004.  In the following section we
introduce the main graphical user interfaces (GUIs) used for the
interactive analysis of IRS data and briefly describe the experienced
user and batch-mode capabilities.

\section{SMART GUIs}

Before starting a SMART analysis session the observers
need to fetch their IRS data from the Spitzer archive.  SMART is designed to
operate on the SSC pipeline basic calibrated data (BCD) FITS files. A
BCD is a calibrated, flatfielded 2-d image. The observer obtains a
BCD image for each IRS exposure (for observing mode and pipeline
details, see SOM7). In addition to the BCD file SMART also needs two
associated files for each exposure. These files are the uncertainty
data and the bad pixel mask. The bad pixel mask has a 16 bit integer
assigned to each pixel. Each bit corresponds to a given warning/error
condition detected during the pipeline processing. For example a pixel
may have suffered a cosmic ray hit or may be saturated.  A perl script
searches the local directory for these 3 files and builds a new FITS
file (`*bcd3p.fits') for each exposure. Each new FITS
file contains the data plane and two extensions: the uncertainty plane
and the bad pixel mask plane.

\subsection{Project Manager}

SSC pipeline products are read into the project manager and either
proceed directly or via image analysis into the ISAP-based Data
Evaluation and Analysis GUI (IDEA). Figure 1 presents a flow chart
outlining the main graphical user interfaces (GUIs) available in
SMART.  At first glance some of the GUIs may appear complex, but we
remind the reader that the SUG will be available at the SMART website.

Figures 2a \& 2b show the project manager and dataset GUIs. These form
the base for launching different applications and storing the
resulting data products. The project manager allows the Spitzer
observer to load files by browsing a local directory containing the
2-d images and spectral table files from the archive or import
files from an existing local data base.  It is designed to handle
large data sets by grouping them into sets called ``projects''.  BCDs
within one project may or may not come from the same IRS module or
astronomical target.  For example, consider the simple case of a
low-resolution observation of a point source covering the full
wavelength range (5.3 $-$ 38 \ums).  This requires the use of both the
SL and LL modules. Each module covers its nominal spectral range in
two orders via two sub-slits. The default observing mode will obtain
two spectra of the target source per sub-slit resulting in eight
separate exposures. The resulting spectra from this observation would
consist of two sets of four spectra from 5.2 - 8.7 \um (SL2), 7.4 -
14.5 \um (SL1), 14.0 - 21.3 \um (LL2) and 19.5 - 38.0 \um (LL2) of the
target source. The IRS low resolution observations always obtain data
simultaneously in the two sub-slits, so there are an additional two
sets of four spectra, with the same wavelength coverage as above, of
the background sky. Entire projects can be saved to disk as IDL save
sets, which can be imported into new SMART projects.
Alternatively individual files can be exported from the project to
disk in either FITS or ascii table format. The main applications
launched from the project manager are described in the following
sections.

\subsection{Image Display/Analysis}

\subsubsection{ ATV-IRS}

We have enhanced a version of the image display program ATV, to work
with our IRS spectral 2-d images. The ATV code was developed to
visualize both 2-d and 3-d images \citep{bar01}. Figure 3 shows an
example of data displayed in ATV-IRS. We have enhanced the code so
that an over-plot tracing the curved spectral orders and the
boundaries of the individual resolution elements can be displayed. The
cursor position is reported in terms of pixel position (x,y) and flux
(pixel value) as well as sky coordinates (right ascension,
declination) and spectral wavelength. This is very useful for
assessing whether weak features in a spectrum are emission lines or
are caused by cosmic-ray hits to the detectors. The viewer can also
display the uncertainty and bad pixel mask planes, returning the same
information for the cursor position. Additional pixels may be flagged
at this stage by editing the bad pixel mask. In addition to displaying
a single BCD, one can make a movie of a stack of images or make a
single mosaiced image.  A table containing the pixel information can
also be inspected. If a stack of images is selected the statistics on
the cube of pixel data can be displayed in a table.  Images from an
external archive can also be added to the project manager and viewed
in ATV-IRS.

\subsubsection{ IDP3-IRS}

The Image Display Paradigm 3 (IDP3) is a sophisticated photometry
software package. It is written in IDL and is designed for the
analysis of the Hubble NICMOS data.  The IRS has imaging capabilities
provided by the two peak-up cameras, each with an $\sim$ 1 arcmin$^2$
field of view. The blue IRS peak-up camera fills a gap in the
wavelength coverage of the Spitzer imagers at 16 \ums.  Both the red
(22 \ums) and blue cameras are used for many observations. Elizabeth
Stobie at the University of Arizona provided a modified version of
IDP3, known as IDP3-IRS, which is optimized for analyzing sources 
observed with the IRS peak-up mode.

\subsubsection{ Quick Look} 

Quick Look takes a BCD file and collapses the spectral orders along
the dispersion direction to produce an average intensity profile
across the source. This provides a convenient tool to search for
either extended emission or weak secondary point sources in the low
resolution data, which have slits that are 57 $''$ (SL) and 168 $''$
(LL) in the cross-dispersion direction. Figure 4 shows an example of a
serendipitous detection of a weak source located close to the slit
center in LL1. The negative stripes in LL2 are caused by the sky
subtraction in LL1 using data which has the target source in LL2, see
Section 2.4 for a discussion of sky subtraction.

\subsubsection{ Image Operations}

The standard image operations - averaging, median filtering, division,
addition and subtraction - are available in SMART.  In addition to
operations weighted by the uncertainty data, pixels can be discarded
according to their bad pixel mask value. This offers a powerful means
for discarding corrupted data and improving the signal to noise. For
example, consider the median of ten BCDs. First the pixels flagged by
a bad pixel mask value are excluded from the calculation of the median
value for each pixel in the 128x128 array. A new bad pixel mask is
generated for the median data using the 'OR' operator on all the bad
pixel mask values associated with the data used to estimate the
median. Automatic co-adding and differencing are available for SL
and LL data.  The images are sorted by slit and position within the slit
(i.e., nod position) and then co-added. The co-added ``on-source'' and
``off-source'' images may then be differenced.

\subsection{Spectral Extraction}

There are currently four extraction routines available in SMART. All
methods use a look-up table supplied by the SSC, which traces the
spectral orders on the BCD image. This table is converted into an IDL
structure known as the ``wavesamp''. The extraction routines use the
wavesamp in order to sub-set the relevant group of pixels in each
resolution element on the array. An example of the wavesamp trace is
shown in Figure 3, where the curved spectral order has been
sub-divided into spectral resolution elements. The curvature results
in fractional pixels being assigned to a given resolution element. The
value of each fractional pixel is scaled by its geometrical area.  The
BCD data is in units of electron/sec.  Each routine estimates the
total number of electron/sec in each individual resolution element.
The final step is to apply the flux calibration (i.e., electron/sec to
Janskys) and stitch the orders together into a single spectrum, using
the pipeline ``fluxcon'' tables.

\subsubsection{ Full Aperture Extraction} 

This is the standard extraction method for SH and LH data. It can also
be used for extended objects that fill the SL and LL apertures. The
pixel values in each resolution element, as defined by the wavesamp,
are summed, while accounting for fractional pixels. Prior to
extraction, pixel values set to NaN (Not a Number) in the pipeline or
flagged by the bad pixel mask are replaced with an average value,
which is estimated from the values of the pixels in the same
resolution element as the bad pixel.

\subsubsection{ Column Extraction}  

For SL and LL observations of point sources the standard extraction
method is column extraction. A column of pixels centered on the point
source is extracted. Figure 5 is an example of the aperture used for a
column extraction in LL2. The column traces the spectral order and its
width in the cross-dispersion direction is scaled with the
instrumental point spread function. The user should over-ride the
default width only after careful consultation of the manual and help
pages ( see \url{http://isc.astro.cornell.edu/smart/}). The user defined
width is scaled with wavelength, and requires additional ``on-the-fly''
calibration, see Section 3.  The pixels in the column in each
resolution element are summed, while accounting for fractional pixels.
Prior to extraction pixel values set to NaN or flagged by bad pixel
mask values are replaced with an average value, which is estimated
from the pixels in the column within the same resolution element as
the bad pixel.

\subsubsection{ Extended Source Extraction } 

A column of pixels centered on the extended source is extracted. The
column traces the spectral order and its width is constant with
wavelength. Again, the pixels in each resolution element within the
column are summed, while accounting for fractional pixels.  Pixel
values set to NaN or flagged by bad pixel mask values are
replaced using the method outlined above for column extraction.

\subsubsection{ Gaussian Extraction} 

Gaussian extraction should only be used with care as it requires
additional on-the-fly calibration, see Section 3. The data from the
pixels in each individual resolution element are collapsed in the
dispersion direction. The resulting 1-d trace is fit with
a Gaussian profile. The Gaussian center and width can be frozen to aid
the extraction of weak sources. Pixel values set to NaN or flagged by 
bad pixel mask values are excluded from the fit.

\subsection{Sky Subtraction}

The BCD images include sky emission and possible detector artifacts,
which can be removed in SMART. The first method is applied before
extraction and the remaining two methods are akin to removing a
baseline before measuring a line flux. If no sky image data are
available a zodiacal model can be subtracted from the spectra in IDEA.

\subsubsection{ Super-Sky Subtraction} 

The sky emission is removed by differencing an ``on-source'' and
``super-sky'' image prior to extraction.  A super-sky image can be
created by co-adding multiple ``off-source'' (i.e., sky) BCDs together.
This can be done in SMART using either the image operations GUI or
using one of the available scripts. This method is applicable to all
four modules. For low resolution data the ``super-sky'' image may 
simply be the median of the ``off-source data'', which is acquired as
part of the standard staring mode observation. For high resolution
data separate sky observations are required.

\subsubsection{ Single-Sky Subtraction}

The sky emission is removed during the extraction process.  An
``off-source'' BCD is used for the sky estimate. This can be done in
two ways.  The first method calculates the median sky pixel value in
each resolution element of the off-source BCD. In the second method,
the pixel values in each resolution element of the off-source BCD are
plotted as a function of cross-dispersion distance from the center of
the slit. A first-order polynomial is fit to the resulting intensity
profile to estimate the sky level. For full aperture extraction the
sky value is scaled with the area of the resolution element.  For
column extraction the sky value is scaled to the area of the column with
in a given resolution element.  For Gaussian extraction the sky value
is scaled to the width of the Gaussian.  The scaled sky value is then
subtracted from each respective summed resolution element, column or
integrated Gaussian profile in the source spectrum.

\subsubsection{Local-Sky Subtraction}

The Single-Sky methods are applied to the data. However in this
instance the sky is calculated from the same BCD that contains the
source data.  The pixels in a given resolution element that are not
part of the source column or Gaussian are used to estimate the sky
value.  An example of a selection of suitable sky regions is shown in
Figure 5.

\subsection{IDEA}

IDEA, the ISAP-based Data Evaluation and Analysis Program, is a
comprehensive 1-D spectral analysis package. The code includes the
inherited ISAP software. ISAP was developed for the analysis of spectral
data from ISO (SWS/LWS/PHT-S/Cam-CVF) and provides a wealth of
routines embedded in an easy-to-use graphical environment.  The
ability to analyse ISO spectra has been preserved so that direct
comparisons may be made between data from the two satellites. Figure
6a shows the IDEA GUI, which has been enhanced to fit the special
needs of IRS data. The applications GUI is shown in Figure 6b.
Processing routines include shifting, zapping low signal-to-noise or
corrupted data, defringing, re-binning, unit conversion, combining
spectra with weighted means or medians, filtering and smoothing.
Analysis routines include line fitting, line identification,
continuum-fitting, synthetic photometry (including the IRAS, ISO and
Spitzer filter profiles), zodiacal light modeling, blackbody fitting,
de-reddening and template fitting routines. The spectra can be
imported/exported as FITS, IDL save sets or ascii tables.

\section{Experienced User and Batch Mode}

SMART is supplied with a default set of calibration files from the
SSC, which can be inspected in the Calibration GUI. However the
experienced user can create her/his own set of calibration files and
import them into SMART via the Calibration GUI. The extraction
routines can be tailored to specific source profiles to maximize the
signal to noise. Intermediate pipeline products, for example, the
un-flatfielded data, can be substituted for the BCD image files. This
can be beneficial for the extraction of faint sources and sources that
have weak features. When un-flatfielded data are used the flux
calibration is performed on-the-fly using a default set of calibration
sources. Both the target and the flux calibrator are extracted using
the same parameters. The extracted calibration source is then used to
flux calibrate the extracted target spectrum. Only an experienced user
should over-ride the default calibration file selection.

SMART is also designed for efficient batch mode processing. Many of
the GUI functions are available in batch mode and we are developing a
suite of scripts for the most commonly used functions. For example,
both full aperture, column and Gaussian extraction are available in
batch mode. The script includes two sky removal options. The first
option is to subtract the sky during extraction. Alternatively, the
sky can be removed using the super-sky method. The data from a given
slit and nod position are co-added and then the co-added
``on-source'' and ``off-source'' images are differenced. A single
spectrum is then extracted from the median filtered, sky-subtracted
image.

\section{Summary}

SMART is currently being tested by the IRS team and the participating
legacy teams. In December 2004 we will have a public release of SMART.
The code will be available at our website,
\url{http://isc.astro.cornell.edu/smart/}, where the IRS observer can
find detailed instructions for downloading and installing SMART. The
site includes a SMART Users Guide and recipes for reducing IRS data.
We plan to update and add new functionality to the code as our
understanding of the IRS data analysis evolves. For example, we are
currently working on an optimized extraction algorithm for both high
and low resolution data.  Updates will be posted at the web site.

\acknowledgments  

We would like to thank the following people, the IRS team and the SSC
for their dedicated work in generating the pipeline processed data and
for ongoing calibration work; the ISAP team for allowing us to inherit
and modify the ISAP code and the referee, Eckhard Sturm, for his swift
endorsement of this paper.

This work is based [in part] on observations made
with the Spitzer Space Telescope, which is operated by the Jet
Propulsion Laboratory, California Institute of Technology under NASA
contract 1407. Support for this work was provided by NASA through
Contract Number 1257184 issued by JPL/Caltech.


%LIST ALL AUTHORS UNLESS 7OR MORE 

%figure 1
\begin{figure}
\figurenum{1}
\includegraphics[scale=.45]{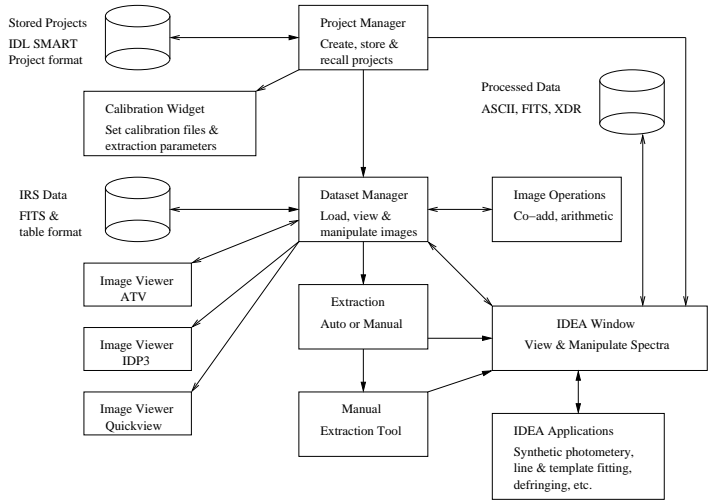}
\caption{SMART Flow Chart}
\end{figure}

%figure 2a
\begin{figure}
\figurenum{2a}
\includegraphics[scale=.45]{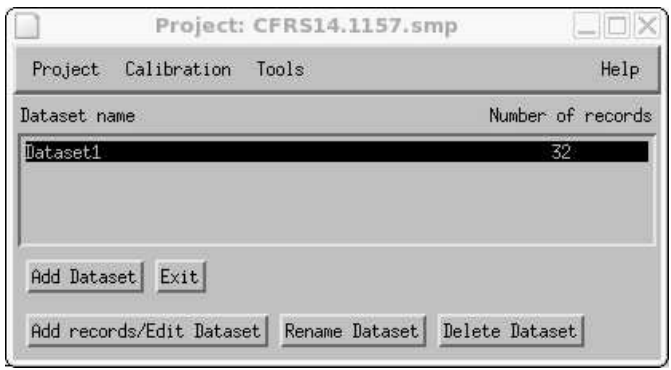}
\caption{The SMART Project Manager GUI}
\end{figure}

%figure 2b
\begin{figure}
\figurenum{2b}
\includegraphics[scale=.45]{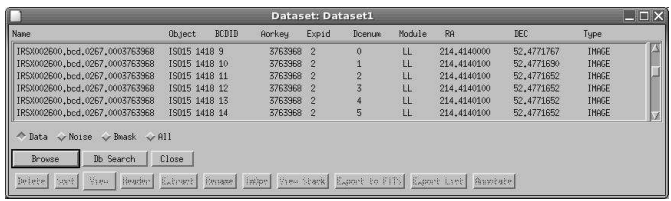}
\caption{The SMART Dataset GUI}
\end{figure}

%figure 3
\begin{figure}
\figurenum{3}
\includegraphics[scale=.45]{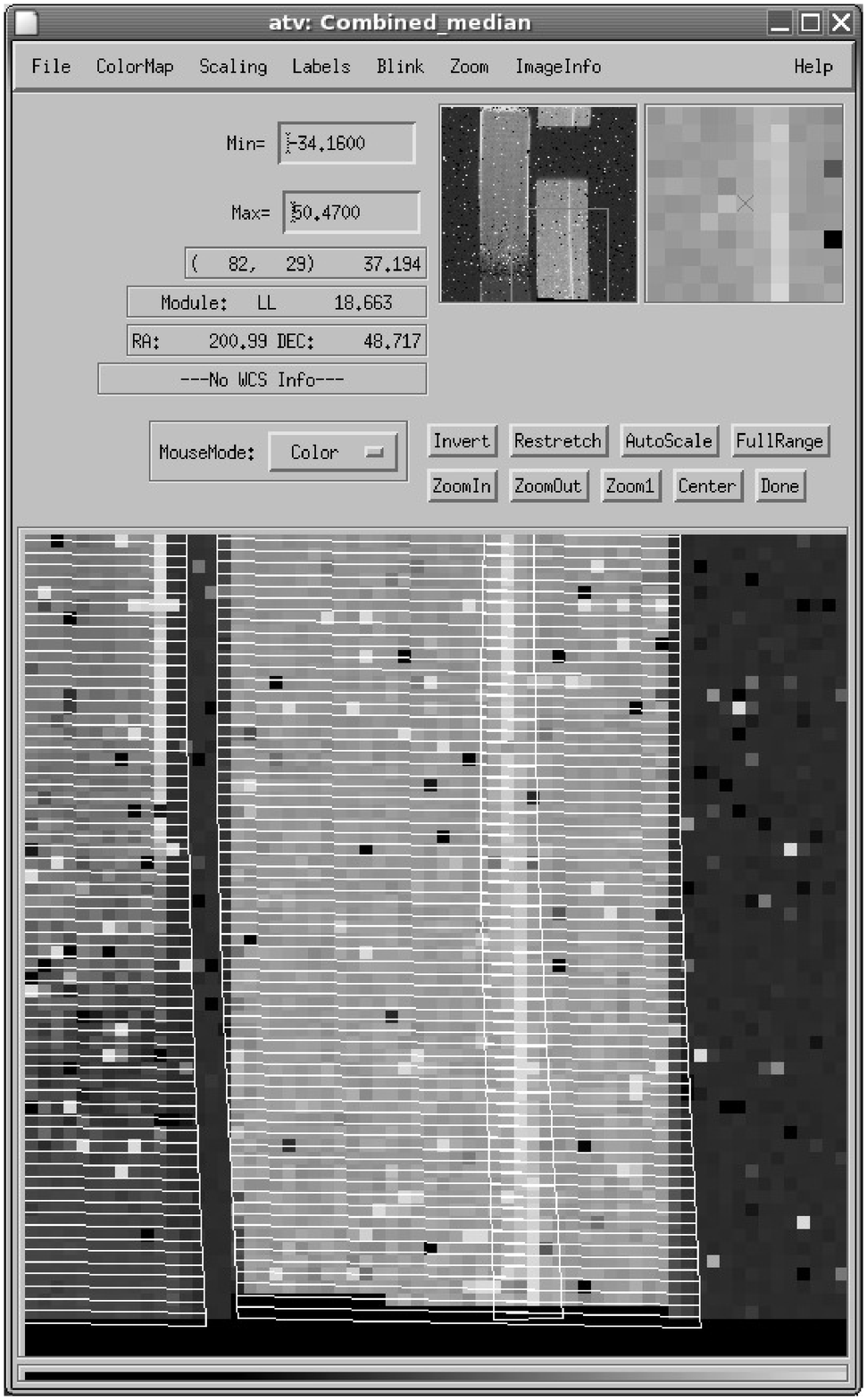}
\caption{The SMART ATV-IRS GUI - Data from IRS LL2 is displayed. Over-plotted are the resolution elements and the column which has been selected for spectral extraction.}
\end{figure}

%figure 4
\begin{figure}
\figurenum{4}
\includegraphics[scale=.45]{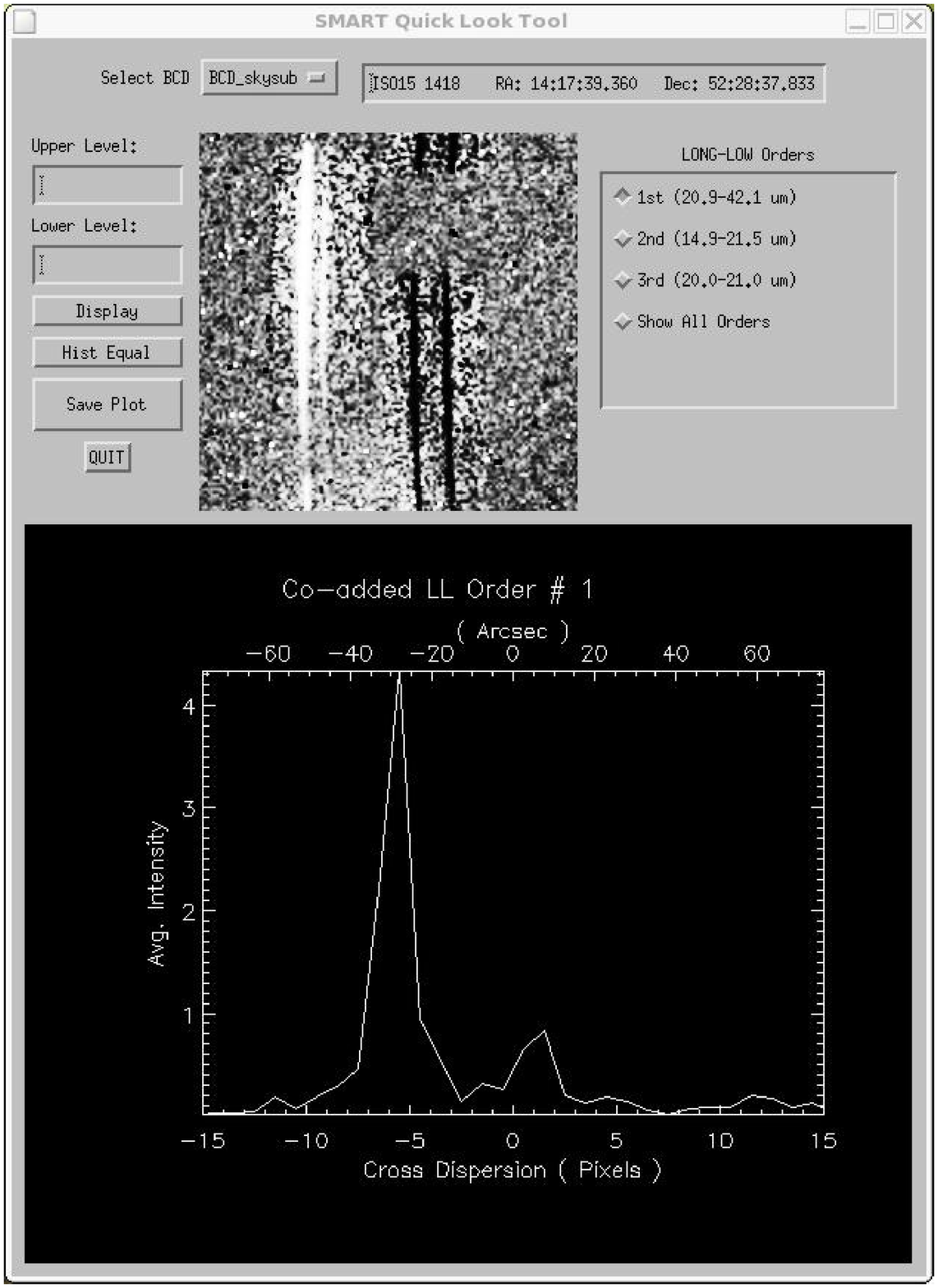}
\caption{The SMART Quick Look GUI - sky subtracted BCD data from the IRS LL is displayed in the upper image window. LL1 has been collapsed in the dispersion direction and is displayed in the plot window. A serendipitous source is observed close to the centre of the slit. The negative stripes in LL2 are caused by the sky
subtraction in LL1 using data which has the target source
in LL2.}
\end{figure}

%figure 5
\begin{figure}
\figurenum{5}
  \includegraphics[scale=.45]{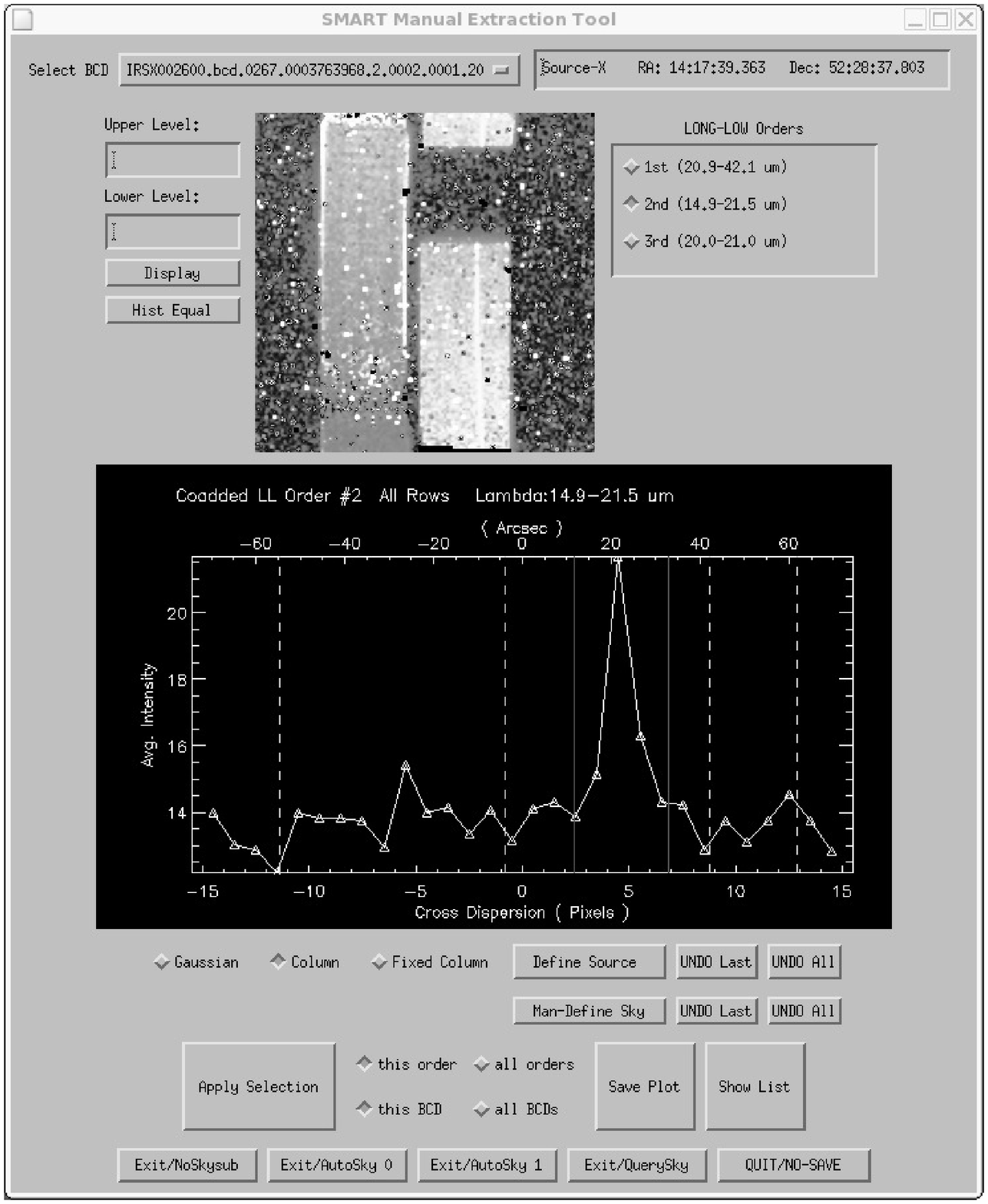}
\caption{The SMART Mandefine GUI - BCD data from the IRS long low spectrometer is displayed in the upper image window. LL2 has been collapsed in the dispersion direction and is displayed in the plot window. Overlaid on this plot are the column boundaries  (dotted-lines) and sky regions (dashed-lines) for column extraction.}
\end{figure}

%figure 6a
\begin{figure}
\figurenum{6a}
\includegraphics[scale=.45]{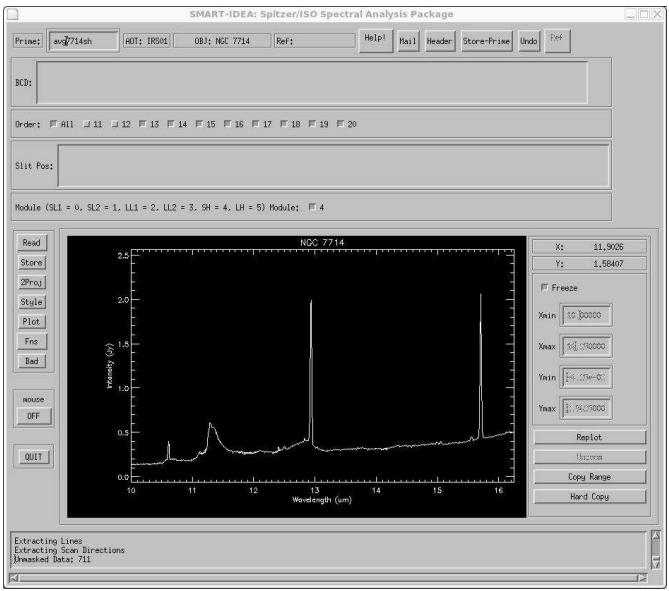}
\caption{The SMART IDEA GUI - this is the main GUI for viewing 1-d spectral data}
\end{figure}

%figure 6b
\begin{figure}
\figurenum{6b}
\includegraphics[scale=.35]{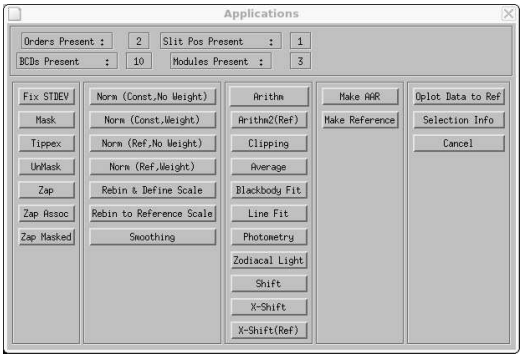}
\caption{The
    SMART IDEA Applications GUI - The majority of the spectral
    reduction and analysis routines are accessed from this GUI}
\end{figure} \end{document}